\documentclass[11pt]{article}

\usepackage[cmtip,arrow]{xy}
\usepackage{pb-diagram,pb-xy}

\newlength{\vshift}
\newlength{\hshift}
\usepackage{amssymb,amsopn}

\def\nn{\nonumber }

\def\be{\beta}
\def\a{\alpha}

\def\ds{\stackrel{\star}{,}}

\def\p{\partial}

\def\p{\partial}

\def\tr{{\rm Tr}}

\def\d{\delta}
\def\G{\Gamma}
\def\S{\Sigma}

\def\nn{\nonumber}
\def\be{\begin{equation}}             \def\ee{\end{equation}}
\def\ba#1{\begin{array}{#1}}          \def\ea{\end{array}}
\def\bea{\begin{eqnarray} }           \def\eea{\end{eqnarray} }
\def\beann{\begin{eqnarray*} }        \def\eeann{\end{eqnarray*} }
\def\beal{\begin{eqalign}}            \def\eeal{\end{eqalign}}
             
\def\bsubeq{\begin{subequations}}     \def\esubeq{\end{subequations}}
\def\bitem{\begin{itemize}}           \def\eitem{\end{itemize}}

\begin{document}

\begin{titlepage}

$\,$

\vspace{1.5cm}
\begin{center}

{\LARGE{\bf (Non)renormalizability of the D-deformed Wess-Zumino model}}

\vspace*{1.3cm}

{{\bf Marija Dimitrijevi\' c${}^{1,2}$, Biljana Nikoli\' c${}^{2}$ and\\
Voja Radovanovi\' c${}^{2}$ }}

\vspace*{1cm}

${}^{1}$ INFN Gruppo collegato di Alessandria\\
Via Bellini 25/G 15100 Alessandria, Italy\\[1em]

${}^{2}$ University of Belgrade, Faculty of Physics\\
Studentski trg 12, 11000 Beograd, Serbia \\[1em]

\end{center}

\vspace*{2cm}

\begin{abstract}

We continue the analysis of the $D$-deformed Wess-Zumino model
which we introduced in the previous paper. The model is defined by a
deformation which is non-hermitian and given in terms of the covariant
derivatives $D_\alpha$. We calculate one-loop divergences in the
two-point, three-point and four-point Green functions. Possibilities
to render the model renormalizable are discussed.

\end{abstract}
\vspace*{1cm}



\vspace*{1cm}
\quad\scriptsize{eMail:
dmarija,biljana,rvoja@ipb.ac.rs}
\vfill

\end{titlepage}\vskip.2cm

\newpage
\setcounter{page}{1}
\newcommand{\Section}[1]{\setcounter{equation}{0}\section{#1}}
\renewcommand{\theequation}{\arabic{section}.\arabic{equation}}

\section{Introduction}

Having in mind problems which physics encounters at small scales
(high energies), in recent years attempts were made to combine
supersymmetry (SUSY) with noncommutative geometry. Different models
were constructed, see for example \cite{luksusy}, \cite{MWsusy},
\cite{nonantisusy}, \cite{Seiberg}, \cite{Ferrara}, \cite{14},
\cite{D-def-us}, \cite{miSUSY}, \cite{Martin08}, \cite{Martin09}.
Some of these models emerge
naturally as low energy limits of string theories in a background
with a constant Neveu-Schwarz two form and/or a constant
Ramond-Ramond two form. For some references on noncommutative
geometry and non(anti)commutative field theories see
\cite{D-def-us}.

One way to describe a noncommutative deformation is to consider
the algebra of functions on a smooth manifold with the usual
pointwise multiplication replaced by a noncommutative
$\star$-product. A wide class of $\star$-products can be defined
by twists ${\cal F}$. The notion of twist was first introduced in
\cite{Drinf} in the context of quantum groups; recently it has
been used to describe symmetries of noncommutative spaces, see for
example \cite{14}, \cite{TwistSymm}, \cite{defgt}.

In our previous paper \cite{D-def-us} we started the analysis of a
simple model. Since we are interested in the renormalizability
properties of the supersymmetric theories with twisted symmetry,
we introduce the non(anti)commutative deformation via the twist
\begin{equation}
{\cal F} = e^{\frac{1}{2}C^{\alpha\beta}D_\alpha \otimes D_\beta
} .\label{intro-twist}
\end{equation}
Here $C^{\alpha\beta} = C^{\beta\alpha}\in \mathbb{C}$ is a
complex constant matrix and $D_\alpha = \partial_\alpha +
i\sigma^m_{\
\alpha\dot{\alpha}}\bar{\theta}^{\dot{\alpha}}\partial_m$ are the
SUSY covariant derivatives. The twist (\ref{intro-twist}) is not
hermitian under the usual complex conjugation. Due to our choice
of the twist, the coproduct of the SUSY transformations remains
undeformed, leading to the undeformed Leibniz rule. The inverse of
(\ref{intro-twist}) defines the $\star$-product. It is obvious
that the $\star$-product of two chiral fields will not be a chiral
field. Therefore we have to use projectors to separate chiral
and antichiral parts. The deformed Wess-Zumino action is then
constructed by inclusion of all possible invariants under the
deformed SUSY transformations.

The plan of the paper is as follows. In the next section we
summarize the most important properties of our model, the details
of the construction are given in \cite{D-def-us}. However, we introduce
two additional terms in the action. These terms are SUSY invariant and they vanish in the commutative limit; they were not considered in the 
previous paper for two reasons. Namely, their presence is not required by the
renormalizability of the two-point function; in addition, they represent
a non-minimal deformation. In other words, they are a deformation of a term
not present in the commutative Wess-Zumino action. We shall see that 
these new terms are essential in order to obtain a renormalizable model. In 
Section 3 we describe the method we are using to calculate divergent
parts of the $n$-point Green functions: the background field method and the supergraph
technique. In Sections 4 and 5 the divergent parts of the two-point, three-point and four-point 
functions are calculated. In Section 6 we discuss renormalizability of the model. In the final section,
we give some comments and compare our results with the results already
present in the literature. Some details of our calculations are
presented in appendix.

\section{$D$-deformed Wess-Zumino model}

We work in the superspace generated by $x^{m}$, $\theta^{\alpha}$
and $\bar{\theta}_{\dot{\alpha}}$ coordinates which fulfill
\begin{eqnarray}
\lbrack x^m, x^n \rbrack = \lbrack x^m,
\theta^\alpha \rbrack = \lbrack x^m, \bar{\theta}_{\dot\alpha} \rbrack = 0 ,\quad
\{ \theta^\alpha , \theta^\beta\} = \{ \bar{\theta}_{\dot\alpha}
, \bar{\theta}_{\dot\beta}\} = \{ \theta^\alpha ,
\bar{\theta}_{\dot\alpha}\} = 0 , \label{undefsupsp}
\end{eqnarray}
with $m=0,\dots 3$ and $\alpha, \beta =1,2$. These coordinates we
call supercoordinates, to $x^m$ we refer as bosonic and to
$\theta^\alpha$ and $\bar{\theta}_{\dot\alpha}$ we refer as
fermionic coordinates. We work in Minkowski space-time with the
metric $(-,+,+,+)$ and $x^2= x^m x_m = -(x^0)^2 + (x^1)^2 + (x^2)^2 +
(x^3)^2$.

A general superfield $F(x,\theta, \bar{\theta})$ can be expanded
in powers of $\theta$ and $\bar{\theta}$,
\begin{eqnarray}
F(x, \theta, \bar{\theta}) &=&\hspace*{-2mm} f(x) + \theta\phi(x)
+ \bar{\theta}\bar{\chi}(x)
+ \theta\theta m(x) + \bar{\theta}\bar{\theta} n(x) + \theta\sigma^m\bar{\theta}v_m(x)\nonumber\\
&& + \theta\theta\bar{\theta}\bar{\lambda}(x) + \bar{\theta}\bar{\theta}\theta\varphi(x)
+ \theta\theta\bar{\theta}\bar{\theta} d(x) .\label{F}
\end{eqnarray}
Under the infinitesimal SUSY transformations it transforms as
\begin{equation}
\delta_\xi F = \big(\xi Q + \bar{\xi}\bar{Q} \big) F, \label{susytr}
\end{equation}
where $\xi^{\alpha}$ and $\bar{\xi}_{\dot{\alpha}}$ are constant
anticommuting parameters and $Q^{\alpha}$ and
$\bar{Q}_{\dot\alpha}$ are SUSY generators,
\begin{eqnarray}
Q_\alpha = \p_\alpha
- i\sigma^m_{\ \alpha\dot{\alpha}}\bar{\theta}^{\dot{\alpha}}\p_m, \quad \bar{Q}_{\dot{\alpha}} = -\bar{\p}_{\dot{\alpha}} + i\theta^\alpha \sigma^m_{\
\alpha\dot{\alpha}}\p_m
.\label{q,barq}
\end{eqnarray}

As in \cite{miSUSY}, \cite{defgt} we introduce a deformation of the
Hopf algebra of infinitesimal SUSY transformations by choosing the
twist $\cal{F}$ in the following way
\begin{equation}
{\cal F} = e^{\frac{1}{2}C^{\alpha\beta}D_\alpha \otimes D_\beta
} ,\label{twist}
\end{equation}
with the complex constant matrix $C^{\alpha\beta} = C^{\beta\alpha}\in
\mathbb{C}$. Note that this twist\footnote{Strictly speaking, the
twist ${\cal F}$ (\ref{twist}) does not belong to the universal
enveloping algebra of the Lie algebra of infinitesimal SUSY
transformations. Therefore to be mathematically correct we should
enlarge the algebra by introducing the relations for the operators
$D_{\alpha}$ as well. In this way the deformed SUSY Hopf algebra
remains the same as the undeformed one. However, since $\lbrack
D_\alpha, M_{mn}\rbrack \neq 0$ the super Poincar\' e algebra
becomes deformed and different from the super Poincar\' e algebra
in the commutative case.} is not hermitian, ${\cal F}^* \neq {\cal
F}$; the usual complex conjugation is denoted by "$*$". It can
be shown that (\ref{twist}) satisfies all requirements for a
twist, \cite{chpr}. The Hopf algebra of infinitesimal SUSY transformations
does not change since
\begin{equation}
\{ Q_\alpha , D_\beta\} = \{ \bar{Q}_{\dot{\alpha}} , D_\beta\} = 0 .\label{komQD}
\end{equation}
This means that the full supersymmetry is preserved.

The inverse of the twist (\ref{twist}),
\begin{equation}
{\cal F}^{-1} = e^{-\frac{1}{2}C^{\alpha\beta}D_\alpha
\otimes D_\beta } ,\label{invtwist}
\end{equation}
defines the $\star$-product. For arbitrary superfields $F$ and $G$
the $\star$-product reads
\begin{eqnarray}
F\star G &=& \mu_\star \{ F\otimes G \} \nonumber\\
&=& \mu \{ {\cal F}^{-1}\, F\otimes G\} \nonumber\\
&=& F\cdot G - \frac{1}{2}(-1)^{|F|}C^{\alpha\beta}(D_\alpha
F)\cdot(D_\beta G) \nonumber\\
&&- \frac{1}{8}C^{\alpha\beta}C^{\gamma\delta}(D_\alpha D_\gamma F)\cdot
(D_\beta D_\delta G)
, \label{star}
\end{eqnarray}
where $|F| = 1$ if $F$ is odd (fermionic) and $|F|=0$ if $F$ is
even (bosonic). The second line is in fact the definition of
the multiplication $\mu_\star$. No higher powers of $C^{\alpha\beta}$
appear since derivatives $D_\alpha$ are Grassmanian. The
$\star$-product (\ref{star}) is associative\footnote{The associativity of the $\star$-product follows from the cocycle condition \cite{chpr} which the twist ${\cal F}$ has to fulfill
\begin{equation}
{\cal F}_{12}(\Delta\otimes id){\cal F} = {\cal F}_{23}(id\otimes\Delta){\cal F}, \label{cocycle}
\end{equation}
where ${\cal F}_{12} = {\cal F}\otimes 1$ and ${\cal F}_{23} =
1\otimes {\cal F}$. It can be shown that the twist (\ref{twist})
indeed fulfills this condition, see \cite{twist2} for details.},
noncommutative and in the zeroth order in the deformation
parameter $C_{\alpha\beta}$ it reduces to the usual pointwise
multiplication. One should also note that it is not hermitian,
\begin{equation}
(F\star G)^* \neq G^* \star F^* . \label{complconj}
\end{equation}

The $\star$-product (\ref{star}) leads to
\begin{eqnarray}
\{ \theta^\alpha \ds \theta^\beta \} &=& C^{\alpha\beta}, \quad \{
\bar{\theta}_{\dot\alpha}\ds \bar{\theta}_{\dot\beta}\} = \{ \theta^\alpha \ds
\bar{\theta}_{\dot\alpha} \} = 0 ,\nonumber \\
\lbrack x^m \ds x^n \rbrack &=& -C^{\alpha\beta}(\sigma^{mn}\varepsilon)_{\alpha\beta}\bar{\theta}\bar{\theta} , \nonumber \\
\lbrack x^m \ds \theta^\alpha \rbrack
&=& -iC^{\alpha\beta}\sigma^m_{\beta\dot{\beta}}\bar{\theta}^{\dot{\beta}},
\quad \lbrack x^m \ds \bar{\theta}_{\dot\alpha} \rbrack = 0 .
\label{thetastar}
\end{eqnarray}
The deformed superspace is generated by the usual bosonic and
fermionic coordinates (\ref{undefsupsp}) while the deformation is
contained in the new product (\ref{star}). From (\ref{thetastar})
it follows that both fermionic and bosonic part of the superspace
are deformed. This is different from \cite{miSUSY} where only
fermionic part of the superspace was deformed.

The deformed infinitesimal SUSY transformation is defined in the
following way
\begin{eqnarray}
\delta^\star_\xi F &=& \big(\xi Q + \bar{\xi}\bar{Q} \big) F .\label{defsusytr}
\end{eqnarray}
Since the coproduct is not deformed, the usual Leibniz rule
follows. The $\star$-product of two superfields is again
a superfield; its transformation law is given by
\begin{eqnarray}
\delta^\star_\xi (F\star G) &=& \big(\xi Q + \bar{\xi}\bar{Q} \big) (F\star G) \nonumber\\
&=& (\delta^\star_\xi F)\star G + F\star (\delta^\star_\xi G) . \label{deftrlaw}
\end{eqnarray}

Being interested in a deformation of the Wess-Zumino model, we
need to analyze properties of the $\star$-products of chiral fields. A chiral field $\Phi$ fulfills $\bar{D}_{\dot{\alpha}}\Phi =0$,
where $\bar{D}_{\dot{\alpha}} = -\bar{\p}_{\dot{\alpha}} -
i\theta^\alpha \sigma^m_{\ \alpha\dot{\alpha}}\p_m$ and
$\bar{D}_{\dot{\alpha}}$ is related to $D_\alpha$ by the usual
complex conjugation. In terms of the component fields, $\Phi$ is
given by
\begin{eqnarray}
\Phi(x, \theta, \bar{\theta}) &=& A(x) + \sqrt{2}\theta^\alpha\psi_\alpha(x)
+ \theta\theta H(x) + i\theta\sigma^l\bar{\theta}(\p_l A(x)) \nonumber\\
&& -\frac{i}{\sqrt{2}}\theta\theta(\p_m\psi^\alpha(x))\sigma^m_{\ \alpha\dot{\alpha}}\bar{\theta}^{\dot{\alpha}}
+ \frac{1}{4}\theta\theta\bar{\theta}\bar{\theta}(\Box A(x)).\label{chiral}
\end{eqnarray}
The $\star$-product of two chiral fields reads
\begin{eqnarray}
\Phi\star\Phi &=& \Phi\cdot\Phi-\frac{1}{8}C^{\alpha\beta}C^{\gamma\delta}D_\alpha
D_\gamma\Phi D_\beta D_\delta \Phi \nonumber\\
&=& \Phi\cdot\Phi -\frac{1}{32}C^2 (D^2\Phi)(D^2\Phi)\nonumber\\
&=& A^2 - \frac{C^2}{2}H^2 + 2\sqrt{2}A\theta^\alpha\psi_\alpha \nonumber\\
&& -i\sqrt{2}C^2 H\bar{\theta}_{\dot{\alpha}}
\bar{\sigma}^{m\dot{\alpha}\alpha}(\p_m\psi_\alpha)
+ \theta\theta \Big( 2AH - \psi\psi\Big) \nonumber\\
&& + C^2\bar{\theta}\bar{\theta}\Big( - H\Box A
+ \frac{1}{2}(\p_m\psi)\sigma^m\bar{\sigma}^l(\p_l\psi) \big) \Big) \nonumber\\
&& + i\theta\sigma^m\bar{\theta}\Big( \p_m (A^2) +C^2H\p_m H \Big) \nonumber\\
&& +
i\sqrt{2}\theta\theta\bar{\theta}_{\dot{\alpha}}\bar{\sigma}^{m\dot{\alpha}\alpha}
\big( \p_m(\psi_\alpha A)\big)\nonumber\\
&&+\frac{\sqrt{2}}{2}\bar\theta\bar\theta C^2(-H\theta\Box  \psi+\theta\sigma^m\bar\sigma^n\p_n \psi\p_m H)\nonumber\\
&&+ \frac{1}{4}\theta\theta\bar{\theta}\bar{\theta} (\Box
A^2-\frac12C^2\Box H^2) , \label{phistarphi}
\end{eqnarray}
where $C^2 =
C^{\alpha\beta}C^{\gamma\delta}\varepsilon_{\alpha\gamma}\varepsilon_{\beta\delta}$.
Due to the $\bar{\theta}$, $\bar{\theta}\bar{\theta}$ and the
$\theta\bar{\theta}\bar{\theta}$ terms in
(\ref{phistarphi}), $\Phi\star\Phi$ is not a chiral field.
Following the method developed in \cite{miSUSY} we decompose all
$\star$-products of the chiral fields into their irreducible
components by using the projectors defined in \cite{wessbook}.

Finally, the deformed Wess-Zumino action is constructed by requiring that 
the action is invariant under the deformed SUSY transformations 
(\ref{defsusytr}) and that in the commutative limit it reduces to the 
undeformed Wess-Zumino action. In addition, we try to make a minimal 
deformation in the sense that we deform by $\star$-multiplication
only the terms already present in the undeformed Wess-Zumino action. 
However, as we shall see latter, renormalizability will in fact imply 
the addition of some 'nonminimal' terms. Thus, we propose the following 
action
\begin{eqnarray}
S &=& \int {\mbox{d}}^4 x\hspace{1mm} \Big\{ \Phi^+\star\Phi\Big|_{\theta\theta\bar\theta\bar\theta}
+ \Big [\frac{m}{2}\Big( P_2(\Phi\star\Phi)\Big|_{\theta\theta} + 2a_1P_1(\Phi\star\Phi)
\Big|_{\bar\theta\bar\theta}\Big) \nn\\
&& +\frac{\lambda}{3}\Big( P_2(P_2(\Phi\star\Phi)\star\Phi)\Big|_{\theta\theta}
+3a_2P_1(P_2(\Phi\star\Phi)\star\Phi)\Big|_{\bar\theta\bar\theta}\nn\\
&& + 2a_3 (P_1(\Phi\star\Phi)\star\Phi)\Big|_{\theta\theta\bar{\theta}\bar{\theta}}
+ 3a_4P_1(\Phi\star\Phi)\star \Phi^+\Big|_{\bar\theta\bar\theta} \nonumber\\
&& + 3a_5 \bar{C}^2P_2(\Phi\star\Phi)\star \Phi^+\Big|_{\theta\theta\bar{\theta}
\bar{\theta}}\Big)
+ {\mbox{ c.c. }} \Big] \Big\}.\label{clas-action}
\end{eqnarray}
Coefficients $a_1,\dots, a_5$ are real and constant. Compared with the action
constructed in \cite{D-def-us}, action (\ref{clas-action}) has two additional terms: the terms with coefficients $a_4$ and $a_5$. Both terms are SUSY invariant
and vanish in the commutative limit. Note that the vanishing
of the $a_5$-term in the commutative limit was done by hand by multiplication with $\bar{C}^2$. These terms were not considered in \cite{D-def-us} because their presence
was not required by renormalizability of the two-point function. In addition,
they are a deformation of a term not present in the commutative
Wess-Zumino action. However, we shall see in the following sections that on the
level of three-point functions, the $a_3$-term generates divergences of the form
$P_1(\Phi\star\Phi)\star \Phi^+\Big|_{\bar\theta\bar\theta}$ while the $a_1$-term and the
$a_4$-term
generate divergences of the form $P_2(\Phi\star\Phi)\star \Phi^+\Big|_{\theta\theta\bar{\theta}
\bar{\theta}}$. In order to absorb these divergences one needs to introduce
the $a_4$-term and the $a_5$-term in the action (\ref{clas-action}) from the very beginning.

\section{The one-loop effective action and the supergraph technique}

In this section we look at the quantum properties of our model. To
be more precise, we calculate the one-loop divergent part of the
effective action up to second order in the deformation parameter.
We use the background field method, dimensional regularization and
the supergraph technique. Note that the use of the supergraph
technique significantly simplifies calculations.

In order to apply the supergraph technique, the classical action
(\ref{clas-action}) has to be rewritten as an integral over the
whole superspace. The kinetic part takes the form
\begin{equation}
S_0 = \int {\mbox{d}}^8 z\hspace{1mm} \Big\{ \Phi^{+}\Phi+
\Big[-\frac{m}{8} \Phi\frac{D^2}{\square}\Phi +\frac{m a_1 C^2}{8}
(D^2\Phi)\Phi + c.c.\Big] \Big\} , \label{S-cl-kin}
\end{equation}
while the interaction is given by
\begin{eqnarray}
S_{int} &=& \lambda\int {\mbox{d}}^8 z\hspace{1mm} \Big\{
-\Phi^2\frac{D^2}{12\square}\Phi + \frac{a_2 C^2}{8}
\Phi\Phi(D^2\Phi)\nonumber \\
&& -\frac{a_3 C^2}{48} (D^2 \Phi)(D^2 \Phi)\Phi
+\frac{a_4 C^2}{8}  \Phi(D^2 \Phi)\Phi^+ \nonumber\\
&& + a_5 \bar{C}^2 \Phi\Phi\Phi^+
+ c.c.\Big\} , \label{S-cl-int}
\end{eqnarray}
with $f(x)\frac{1}{\Box}g(x) = f(x)\int {\mbox{d}}^4 y\hspace{1mm} G(x-y)g(y)$. Following the
idea of the background field method, we split the chiral and antichiral superfields into their classical and quantum parts
\begin{eqnarray}
\Phi\to \Phi+ \Phi_q ,\quad \Phi^+ \to \Phi^{+}+\Phi^{+}_q
\end{eqnarray}
and integrate over the quantum superfields in the path integral.

Since $\Phi_q$ and $\Phi_q^{+}$ are chiral and antichiral fields, they are constrained by
$$\bar D_{\dot\a}\Phi_q= D_{\a}\Phi_q^+=0 .$$
One can introduce unconstrained superfields $\S$ and $\S^+$ such
that
\begin{eqnarray}
\Phi_{q}&=&-\frac{1}{4}\bar{D}^2\Sigma \ ,\nonumber \\
\Phi_{q}^{+}&=&-\frac{1}{4}D^2\Sigma^{+}\ .\label{abel-gaug-tr}
\end{eqnarray}
Note that we do not express the background superfields $\Phi$ and
$\Phi^+$ in terms of $\S$ and $\S^+$, only the quantum parts
$\Phi_q$ and $\Phi_q^+$. After integrating over the quantum
superfields, the result will be expressed in terms of the
(anti)chiral superfields. This is a big advantage of the
background field method (and the supergraph technique). From
(\ref{abel-gaug-tr}) we see that the unconstrained superfields are
determined up to a gauge transformation
\begin{eqnarray}
\S&\to&\S+\bar
D_{\dot\a}\bar\Lambda^{\dot\a} \ , \nonumber\\
\S^+&\to&\S^++D^{\a}\Lambda_{\a} \ ,
\end{eqnarray}
with the gauge parameter $\Lambda$. In order to fix this
symmetry we have to add a gauge fixing term to the action. For the
gauge functions we choose
\begin{eqnarray}
\chi_\a&=&D_\a\S \ , \nonumber\\
\bar \chi_{\dot\a}&=&\bar
D_{\dot\a}\S^+ \ .
\end{eqnarray}
The product $\d(\chi)\d(\bar\chi)$ in the path integral is
averaged by the weight $e^{-i\xi \int d^8z\bar{f}Mf}$:
\begin{equation}
\int {\mbox{d}}f {\mbox{d}}\bar f \hspace{1mm}
\d(\chi_\a-f_\a)\d(\bar\chi^{\dot\a}-\bar f^{\dot\a})e^{-i\xi\int
d^8 z\bar f^{\dot\a}M_{\dot\a \a}f^\a}
\end{equation}
where
\begin{equation}
\bar f^{\dot \a}M_{\dot\a\a}f^\a=\frac{1}{4}\bar f^{\dot
\a}(D_\a\bar D_{\dot\a}+\frac{3}{4}\bar D_{\dot\a}D_\a)f^\a
\end{equation}
and the gauge fixing parameter is denoted by $\xi$. The gauge
fixing term becomes
\begin{equation}
S_{gf}=-\xi\int
{\mbox{d}}^8 z\hspace{1mm} (\bar D_{\dot\a}\bar\S)(\frac{3}{16}\bar
D^{\dot\a}D^{\a}+\frac 14 D^\a\bar D^{\dot\a})(D_\a\S) .
\end{equation}
It is easy to show that the ghost fields are decoupled.

The part of the classical gauge fixed action quadratic in quantum
superfields is
\begin{equation}
S^{(2)}=\frac{1}{2}\int {\mbox{d}}^8 z\hspace{1mm} \left(\begin{array}{cc}
\Sigma & \Sigma^{+} \end{array} \right) \Big({\mathcal
M}+{\mathcal V}\Big) \left(\begin{array}{c}\Sigma\\
\Sigma^{+} \end{array} \right)
\end{equation}
%
where the kinetic and the interaction terms are collected in
the matrices ${\mathcal M}$ and ${\mathcal V}$ respectively. Matrix
${\mathcal M}$ is given by
\begin{equation}
\mathcal M= \left(\begin{array}{cc} -m\square^{1/2}(1-a_1 C^2
\square)P_{-} & \square (P_2+\xi(P_1+P_T))\\
\square (P_1+\xi(P_2+P_T)) & -m\square^{1/2}(1-a_1 \bar{C}^2 \square)P_{+}
\end{array} \right) .
\end{equation}
Matrix ${\mathcal V}$ has the form
\begin{equation}
{\mathcal V}=\left(\begin{array}{cc} F & G\\
\bar{G} & \bar{F}\end{array}\right)\ ,
\end{equation}
with matrix elements
\begin{eqnarray}
F&=&-\frac{\lambda}{2} \Phi \bar{D}^2 + \frac{\lambda a_2 C^2}{2}
\Phi\square\bar{D}^2 + \frac{\lambda a_2
C^2}{4}(\square \Phi)\bar{D}^2\nonumber \\
&&- \frac{\lambda a_3 C^2}{128}\overleftarrow{\bar{D}^2}
(D^2\Phi)D^2\bar{D}^2 + \frac{\lambda a_4 C^2}{64}
\overleftarrow{\bar{D}^2}
(\Phi^{+})D^2\bar{D}^2 \nonumber\\
&& +\frac{\lambda a_5}{8}\bar{C}^2 (\bar{D}^2\Phi^+) 
\bar{D}^2, \nonumber\\
G &=& \frac{\lambda a_4}{64}\overleftarrow{\bar{D}^2}
\Big[ C^2(D^2\Phi) +\bar{C}^2(\bar{D}^2\Phi^{+})\Big] D^2 \nonumber\\
&& + \frac{\lambda a_5}{8}
\overleftarrow{\bar{D}^2} \Big[ C^2\Phi^+ + \bar{C}^2\Phi \Big] D^2\ ,
\nonumber \\
\bar{F}&=&-\frac{\lambda}{2} \Phi^{+} D^2 + \frac{\lambda a_2
\bar{C}^2}{2} \Phi^{+}\square D^2 + \frac{\lambda a_2
\bar{C}^2}{4}(\square \Phi^{+})D^2\nonumber \\
&&- \frac{\lambda a_3 \bar{C}^2}{128}\overleftarrow{D^2}
(\bar{D}^2\Phi^{+})\bar{D}^2 D^2 + \frac{\lambda a_4
\bar{C}^2}{64} \overleftarrow{D^2}
(\Phi)\bar{D}^2 D^2 \nonumber\\
&& + \frac{\lambda a_5}{8} C^2 (D^2 \Phi) D^2, \nonumber\\
\bar{G} &=& \frac{\lambda a_4}{64}\overleftarrow{D^2}
\Big[ \bar{C}^2 (\bar{D}^2\Phi^{+}) +C^2(D^2\Phi) \Big] \bar{D}^2 \nonumber\\
&&+ \frac{\lambda a_5}{8}\overleftarrow{D^2} \Big[ \bar{C}^2\Phi + C^2\Phi^+ 
\Big] \bar{D}^2\ .
\end{eqnarray}

The one-loop effective action is
\begin{equation}
\G=S_0+S_{int}+\frac{i}{2}\tr\log(1+{\mathcal M}^{-1}{\mathcal V})
\ .\label{ED-opsta}
\end{equation}
The last term in (\ref{ED-opsta}) is the one-loop correction to
the effective action. In order to calculate it we have to invert
${\mathcal M}$ \cite{wessbook},
\begin{equation}
{\mathcal M}^{-1}=\left(\begin{array}{cc} A & B\\ \bar{B} & \bar{A}
\end{array}\right)=\left(\begin{array}{cc}
\frac{m(1-a_1\bar{C}^2\square)D^2}{4\square f(\square)} &
\frac{D^2\bar{D}^2}{16\square f(\square)}+\frac{\bar{D}^2 D^2-2\bar{D}D^2\bar{D}}{16\xi\square^2}\\
\frac{\bar{D}^2 D^2}{16\square
f(\square)}+\frac{D^2\bar{D}^2-2D\bar{D}^2 D}{16\xi\square^2} &
\frac{m(1-a_1 C^2\square)\bar{D}^2}{4\square f(\square)}
\end{array}\right)
\end{equation}
where
\begin{equation}
f(\square)=\square-m^2+m^2 a_1 (C^2+\bar{C}^2)\square .
\end{equation}
Expansion of the logarithm in (\ref{ED-opsta}) gives the one-loop
correction to the effective action
\begin{equation}
\Gamma_1=\frac{i }{2}\tr\sum_{n=1}^{\infty}
\frac{(-1)^{n+1}}{n}({\mathcal
M}^{-1}\mathcal{V})^{n}=\sum^{\infty}_{n=1} \Gamma_{1}^{(n)}.
\label{1LoopExp}
\end{equation}

\section{Two-point and three-point Green functions}

Let us now look at the divergent parts of the effective action.
The first term in the expansion (\ref{1LoopExp}) is given by
\begin{equation}
\Gamma^{(1)}_{1}=\frac{i}{2} \tr ({\mathcal M}^{-1}{\mathcal
V})=\frac{i}{2} \tr \Big[AF+ \bar{A}\bar{F}+ B\bar{G} + \bar{B}G
\Big] = 0 \ . \label{tadpole}
\end{equation}
This means that the tadpole contribution to the one-loop effective
action vanishes just as in the undeformed Wess-Zumino model.

The second term in (\ref{1LoopExp}) contains two classical fields
and gives the one-loop divergent part of the two-point function.
It is given by
\begin{eqnarray}
\Gamma^{(2)}_{1} &=& -\frac{i}{4} \tr ({\mathcal
M}^{-1}{\mathcal V})^{2} \nonumber\\
&=& -\frac{i}{4} \tr \Big[ AFAF + \bar{A}\bar{F}\bar{A}\bar{F}
+2AFB\bar{G}+ 2\bar{A}\bar{F}\bar{B}G\nonumber\\ & &+ 2AG\bar{B}F+
+2\bar{A}\bar{G}B\bar{F}+ 2 B\bar{F}\bar{B}F \Big] \ .\label{G2}
\end{eqnarray}
In our previous paper \cite{D-def-us} we obtained the result for
the divergent part of the two-point function applying the
background field method to the action written in terms of the
component fields and assuming that $C^2 = \bar{C}^2$. Here we will
not make this assumption. Terms in the classical action (\ref{S-cl-kin})-(\ref{S-cl-int}) which are proportional to $a_2$ give divergences which do not have their counterparts in the classical action and would lead to a nonrenormalizable two-point function. Therefore, we set $a_2 =0$. Note that this is in agreement with \cite{D-def-us}.

The divergent parts of terms appearing in (\ref{G2}) are given by
\begin{eqnarray}
\tr (AFAF)\bigg |_{dp} &=& \frac{i m^2 \lambda^2
C^2}{4\pi^{2}\varepsilon}\int {\mbox{d}}^8 z\hspace{1mm}
\Phi(z)\Big[ a_3
D^2\Phi(z) - 2 a_4\Phi^{+}(z) \Big] ,\label{AFAF}\\
\tr (\bar{A}\bar{F}\bar{A}\bar{F})\bigg |_{dp} &=& \frac{i m^2
\lambda^2 \bar{C}^2}{4\pi^{2}\varepsilon}\int {\mbox{d}}^8
z\hspace{1mm} \Phi^{+}(z)\Big[ a_3
\bar{D}^2\Phi^{+}(z) - 2 a_4\Phi(z) \Big] ,\label{barAbarFbarAbarF}\\
\tr (AFB\bar{G})\bigg |_{dp} &=& \tr
(AG\bar{B}F)\bigg |_{dp} \nonumber\\
&=& -\frac{i m \lambda^2 C^2
a_4}{16\pi^{2}\varepsilon}\int {\mbox{d}}^8 z\hspace{1mm} \Phi(z) D^2\Phi(z) \nonumber\\
&&-\frac{i m \lambda^2 C^2
a_5}{2\pi^{2}\varepsilon}\int {\mbox{d}}^8 z\hspace{1mm} \Phi^+(z) \Phi(z) , 
\label{AFGB} \\
\tr (\bar{A}\bar{F}\bar{B}G)\bigg |_{dp} &=& \tr
(\bar{A}\bar{G}B\bar{F})\bigg |_{dp} \nonumber\\
&=& -\frac{i m \lambda^2 \bar{C}^2
a_4}{16\pi^{2}\varepsilon}\int {\mbox{d}}^8 z\hspace{1mm} \Phi^{+}(z) 
\bar{D}^2\Phi^{+}(z) \nonumber\\
&& -\frac{i m \lambda^2 \bar{C}^2
a_5}{2\pi^{2}\varepsilon}\int {\mbox{d}}^8 z\hspace{1mm} \Phi^{+}(z) \Phi(z), 
\label{barAbarFbarBG} \\
\tr (B\bar{F}\bar{B}F)\bigg |_{dp} &=& \frac{i \lambda^2(1- 2a_1
m^2(C^2+\bar{C}^2))}{2\pi^{2}\varepsilon}\int {\mbox{d}}^8
z\hspace{1mm} \Phi^{+}(z) \Phi(z)\nonumber\\
&& -\frac{i \lambda^2 
a_5}{8\pi^{2}\varepsilon}\int {\mbox{d}}^8 z\hspace{1mm} \Big[ C^2\Phi(D^2\Phi)
+ \bar{C}^2\Phi^+ (\bar{D}^2\Phi^+ \Big] .\label{BFBF}
\end{eqnarray}
From (\ref{AFAF}-\ref{BFBF}) and (\ref{G2}) we obtain
\begin{eqnarray}
\Gamma^{(2)}_{1}\bigg|_{dp}&=&
\frac{\lambda^2(1-(2a_1+\frac{a_4}{2} + \frac{2a_5}{m})m^2(C^2+\bar{C}^2))}{4\pi^2\varepsilon}\int
{\mbox{d}}^8 z\hspace{1mm} \Phi^{+}(z)\Phi(z)\nonumber\\
&& +\frac{\lambda^2 (m^2a_3- ma_4 -a_5)}{16\pi^2\varepsilon}\int
{\mbox{d}}^8 z\hspace{1mm} \left [ C^2\Phi(z)D^2\Phi(z) + c.c.\right
] . \label{G2expl}
\end{eqnarray}
This result can be easily rewritten in terms of the component
fields. Choosing $a_4 = a_5=0$ one sees that (\ref{G2expl}) is in agreement 
with the results obtained in \cite{D-def-us}.

Let us next consider the divergent part of the three-point Green
function
\begin{eqnarray}
\Gamma^{(3)}_{1}&=&\frac{i}{6} \tr ({\mathcal
M}^{-1}{\mathcal V})^{3} \nonumber\\
&=& \frac{i}{6} \tr \Big[ AFAFAF +
\bar{A}\bar{F}\bar{A}\bar{F}\bar{A}\bar{F} +3AFAFB\bar{G}
+3\bar{A}\bar{F}\bar{A}\bar{F}\bar{B}G+ 3AFAG\bar{B}F\nonumber \\
& & + 3\bar{A}\bar{F}\bar{A}\bar{G}B\bar{F} + 3 AFB\bar{F}\bar{B}F
+ 3 \bar{A}\bar{F}\bar{B}FB\bar{F}+ 3
AFB\bar{F}\bar{A}\bar{G}\nonumber \\ & &+ 3
\bar{A}\bar{F}\bar{B}FAG+ 3B\bar{G}B\bar{F}\bar{B}F +
3\bar{B}G\bar{B}FB\bar{F} \Big] .\label{G3}
\end{eqnarray}
The traces appearing in (\ref{G3}) are equal to
\begin{eqnarray}
\tr (AFB\bar{F}\bar{B}F)\bigg |_{dp}&=&\frac{i m\lambda^3 a_1
\bar{C}^2}{\pi^2\varepsilon}\int {\mbox{d}}^8 z\hspace{1mm}
\Phi(z)\Phi(z)\Phi^{+}(z) \nonumber\\
&& - \frac{i m\lambda^3 a_3 C^2}{2\pi^2\varepsilon}\int
{\mbox{d}}^8 z\hspace{1mm}
\Phi(z)\Phi^{+}(z) D^2\Phi(z) \nonumber\\
&& + \frac{i m\lambda^3 a_4 C^2}{\pi^2\varepsilon}\int
{\mbox{d}}^8 z\hspace{1mm}
\Phi(z)\Phi^{+}(z)\Phi^{+}(z) ,\label{TrG3'}\\
\tr (\bar{A}\bar{F}\bar{B}FB\bar{F})\bigg |_{dp}&=&\frac{i
m\lambda^3 a_1 C^2}{\pi^2\varepsilon}\int {\mbox{d}}^8
z\hspace{1mm}
\Phi^{+}(z)\Phi^{+}(z)\Phi(z) \nonumber\\
&& - \frac{i m\lambda^3 a_3 \bar{C}^2}{2\pi^2\varepsilon}\int
{\mbox{d}}^8 z\hspace{1mm}
\Phi^{+}(z)\Phi(z) \bar{D}^2\Phi^{+}(z) \nonumber\\
&& + \frac{i m\lambda^3 a_4 \bar{C}^2}{\pi^2\varepsilon}\int
{\mbox{d}}^8 z\hspace{1mm}
\Phi^{+}(z)\Phi(z)\Phi(z) ,\label{TrG3'''}\\
\tr (B\bar{G}B\bar{F}\bar{B}F)\bigg |_{dp} &=& \tr
(\bar{B}G\bar{B}FB\bar{F})\bigg |_{dp}\nonumber\\
&=&\frac{i \lambda^3 a_4}{8\pi^2\varepsilon}\int {\mbox{d}}^8
z\hspace{1mm}\Big[ C^2\Phi(z)\Phi^{+}(z)D^2\Phi(z) +c.c. \Big] \nonumber\\
&& + \frac{i m\lambda^3 a_5}{\pi^2\varepsilon}\int
{\mbox{d}}^8 z\hspace{1mm} \Big[ \bar{C}^2\Phi^{+}(z)\Phi(z)\Phi(z) + c.c. \Big]. \label{TrG3''}
\end{eqnarray}
All other terms in (\ref{G3}) are convergent. We obtain
\begin{eqnarray}
\Gamma^{(3)}_{1}\bigg|_{dp}&=& -\frac{
\lambda^3(ma_1+ ma_4 +2a_5)}{2\pi^2\varepsilon}\int
{\mbox{d}}^8 z\hspace{1mm} \left [ \bar{C}^2\Phi(z)\Phi(z)\Phi^{+}(z)+c.c.\right ] \nonumber\\
&& + \frac{\lambda^3 (2ma_3- a_4)}{8\pi^2\varepsilon}\int
{\mbox{d}}^8 z\hspace{1mm} \left [ C^2\Phi(z)\Phi^{+}(z)D^2\Phi(z) +
c.c.\right ] .\label{G3-rezultat}
\end{eqnarray}
From (\ref{G3-rezultat}) we see that the $a_3$-term generates divergences of the
type 
$$\int {\mbox{d}}^8 z\hspace{1mm} \left [ C^2\Phi(z)\Phi^{+}(z)D^2\Phi(z) +
c.c.\right ]$$ 
while the $a_1$-term and the $a_4$-term generate divergence of
the type 
$$\int {\mbox{d}}^8 z\hspace{1mm} \left [ 
\bar{C}^2\Phi(z)\Phi(z)\Phi^{+}(z)+c.c.\right ].$$ 
These divergences cannot be canceled unless we introduce two additional terms in the action (\ref{clas-action}).

\section{Four-point Green function}

Before making the final statement about renormalizability we still have to 
check whether divergences in the four-point function can be canceled. The four-point function is given by
\begin{eqnarray}
\Gamma^{(4)}_{1}&=&-\frac{i}{8} \tr ({\mathcal M}^{-1}{\mathcal V})^{4} \nonumber\\
&=&-\frac{i}{8} \tr \Big[ AFAFAFAF+
\bar{A}\bar{F}\bar{A}\bar{F}\bar{A}\bar{F}\bar{A}\bar{F}+
2B\bar{F}\bar{B}FB\bar{F}\bar{B}F\Big]\nonumber\\ && -\frac{i}{2}
\tr \Big[AFAFB\bar{F}\bar{B}F  +
\bar{A}\bar{F}\bar{A}\bar{F}\bar{B}FB\bar{F}+
AFB\bar{F}\bar{A}\bar{F}\bar{B}F  + AFAFAFB\bar{G}
 \nonumber\\ && + \bar{A}\bar{F}\bar{A}\bar{F}\bar{A}\bar{F}\bar{B}G+
AFAFB\bar{F}\bar{A}\bar{G} +
\bar{A}\bar{F}\bar{A}\bar{F}\bar{B}FAG +AFAFAG\bar{B}F\nonumber\\
&& + \bar{A}\bar{F}\bar{A}\bar{F}\bar{A}\bar{G}B\bar{F}+
AFB\bar{G}B\bar{F}\bar{B}F +
\bar{A}\bar{F}\bar{B}G\bar{B}FB\bar{F} + AFB\bar{F}\bar{B}FB\bar{G}\nonumber\\
&& + \bar{A}\bar{F}\bar{B}FB\bar{F}\bar{B}G+
AFB\bar{F}\bar{A}\bar{F}\bar{A}\bar{G} +
\bar{A}\bar{F}\bar{B}FAFAG + AFB\bar{F}\bar{B}G\bar{B}F\nonumber\\
&& +\bar{A}\bar{F}\bar{B}FB\bar{G}B\bar{F}+
B\bar{F}\bar{B}FB\bar{F}\bar{A}\bar{G} +
\bar{B}FB\bar{F}\bar{B}FAG\Big]\ .\label{G4}
\end{eqnarray}
There is only one non-vanishing divergent term in (\ref{G4}) and it is given by
\begin{eqnarray}
\tr (B\bar{F}\bar{B}FB\bar{F}\bar{B}F)\bigg
|_{dp} &=& \frac{i \lambda^4}{\pi^2\varepsilon}\int {\mbox{d}}^8 z\hspace{1mm}\Big[ \bar{C}^2\Phi(z)\Phi(z)\Phi^{+}(z) \Big( a_3
\bar{D}^2\Phi^{+}(z) \nonumber\\
&& \hspace{3cm} -2a_4 \Phi(z)\Big) + c.c.
\Big ] . \label{TrG4}
\end{eqnarray}
Therefore the divergent part of the four-point function does not
vanish and it is given by
\begin{equation}
\Gamma^{(4)}_{1}\bigg |_{dp}=\frac{ \lambda^4
}{8\pi^2\varepsilon}\int {\mbox{d}}^8 z\hspace{1mm}\left [ \bar
C^2\Phi(z)\Phi(z)\Phi^{+}(z)\left(a_3
\bar{D}^2\Phi^{+}(z)-2a_4\Phi(z)\right) +c.c. \right ]
.\label{G4-rezultat}
\end{equation}
We see that this term does not appear in the classical action
(\ref{clas-action}). One can check that the five-point function is convergent.

Let us note that in the undeformed Wess-Zumino 
model divergences appear only in the two-point function. 
They lead to renormalization of the superfield and there is no mass 
counterterm. Three-point and higher-point functions are convergent, which 
means that there are no divergent counterterms for the coupling constants; all 
redefinitions can be expressed in terms of the field strength renormalization 
$Z$. We see that introducing the deformation (\ref{twist}) changes this 
behavior: we obtain divergences both in the three-point and in the four-point 
functions.

\section{Discussion}

From (\ref{G2expl}), (\ref{G3-rezultat}) and (\ref{G4-rezultat}) we see that 
the two-point and the three-point functions are renormalizable, while the 
four-point function is not. Thus, the model with arbitrary coefficients 
$a_1, \dots, a_5$ is not renormalizable. The situation is very similar to 
the one in the noncommutative gauge theory with the massless Dirac fermion 
$\psi$ \cite{MV}, where only two-point Green functions were
renormalizable. In that model the four-$\psi$ divergence appears
and it cannot be removed.

However, there is a special choice of the coefficients $a_1, \dots a_5$ which 
renders the model renormalizable. If we fix $a_3 = a_4 =0$, the divergent 
part of the four-point function vanishes. In that case the divergent parts 
of the two- and three-point functions are
\begin{eqnarray}
\Gamma^{(2)}_{1}\bigg|_{dp}&=&
\frac{\lambda^2(1-(2a_1 + \frac{2a_5}{m})m^2(C^2+\bar{C}^2))}{4\pi^2\varepsilon}\int
{\mbox{d}}^8 z\hspace{1mm} \Phi^{+}(z)\Phi(z)\nonumber\\
&& -\frac{\lambda^2 a_5}{16\pi^2\varepsilon}\int
{\mbox{d}}^8 z\hspace{1mm} \left [ C^2\Phi(z)D^2\Phi(z) + c.c.\right
] , \label{G2new}\\
\Gamma^{(3)}_{1}\bigg|_{dp}&=& -\frac{
\lambda^3(ma_1 +2a_5)}{2\pi^2\varepsilon}\int
{\mbox{d}}^8 z\hspace{1mm} \left [ 
\bar{C}^2\Phi(z)\Phi(z)\Phi^{+}(z)+c.c.\right ] 
.\label{G3new}
\end{eqnarray}
All divergences in (\ref{G2new}) and (\ref{G3new}) 
have the same form as terms in the classical action 
(\ref{S-cl-kin})-(\ref{S-cl-int}). 
But this is only a necessary condition for a theory to be 
renormalizable; we still have to check the consistency of the field and 
the coupling constants redefinitions. As usual, we add 
counterterms to the
classical action (\ref{S-cl-kin})-(\ref{S-cl-int}). The bare
action is given by
\begin{equation}
S_B=S_0+S_{int}-\Gamma^{(2)}_{1}\bigg |_{dp}-\Gamma^{(3)}_{1}\bigg |_{dp}\ .\label{golo}
\end{equation}
The two-point Green function in (\ref{golo}) gives the
renormalization of the superfield $\Phi$
\begin{equation}
\Phi_0=\sqrt{Z}\Phi\ ,
\end{equation}
where
\begin{equation}
Z=1-\frac{\lambda^2}{4\pi^2\epsilon}\Big[ 1
- 2m(C^2+\bar C^2) (ma_1 + a_5)\Big]
\end{equation}
and
\begin{equation}
m=Zm_0
\end{equation}
since $\d_m=0$. In addition to the field redefinition we obtain the
redefinition of the coupling constant
\begin{equation}
a_{10}C_0^2=a_1C^2\Big[ 
1 + \frac{\lambda^2 a_5}{2\pi^2\epsilon a_1m}\Big]\ .\label{a_1ren}
\end{equation}
From the three-point Green function in (\ref{golo}) we obtain the following 
conditions
\begin{eqnarray}
\lambda&=&Z^{3/2}\lambda_0\ ,\label{newLambda}\\
a_{50}\bar{C}_0^2&=&a_5\bar{C}^2\Big[ 
1+\frac{\lambda^2(ma_1 + 2a_5)}{2\pi^2\epsilon a_5}\Big]  .\label{a_5ren}
\end{eqnarray}
Taking these results into account we see that indeed for $a_3=a_4=0$ our model is
renormalizable.

Let us note that for the special cases $a_5=-\frac{1}{2}ma_1$ and $a_5=2ma_1$ equations (\ref{a_1ren}) and (\ref{a_5ren}) can be reduced to
\begin{equation}
a_{10}C_0^2=a_1C^2\Big[ 
1 - \frac{\lambda^2}{4\pi^2\epsilon }\Big]\ .\label{newa_1ren}
\end{equation}
The case $a_5=-\frac{1}{2}ma_1$ is more interesting. With this choice the divergent part of the three-point function (\ref{G3new}) vanishes. This means that there are no divergent counterterms for the coupling constants, i.e. all redefinitions are expressed in terms of the field strength renormalization
\begin{eqnarray}
m &=& Zm_0, \nonumber\\
\lambda &=& Z^{3/2}\lambda_0 ,\nonumber\\
a_{10}C_0^2 &=& a_1C^2Z .\label{redfNew}
\end{eqnarray}
These results resemble those which are valid for the undeformed Wess-Zumino model.

\section{Conclusions}

In order to see how a deformation by twist affects renormalizability of the Wess-Zumino model, we consider a special
example of the twist, (\ref{twist}). Compared with the classical
SUSY Hopf algebra, the twisted SUSY Hopf algebra is unchanged. In
particular, the Leibniz rule (\ref{deftrlaw}) remains undeformed.
The notion of chirality is however lost and we have to apply the
method of projectors introduced in \cite{miSUSY} to obtain the
action. A deformation of the commutative Wess-Zumino action which is SUSY 
invariant and has a good commutative limit is introduced and its 
renormalizability properties are discussed.

Using the background field method and the supergraph technique we
calculate the divergent parts of the two-point, three-point and
four-point functions up to second order in the deformation
parameter $C_{\alpha\beta}$. For one-point and two-point functions
the obtained results are in agreement with the results of
\cite{D-def-us}, where the analysis was done in component fields.
There is no tadpole diagram, no mass renormalization and all
fields are renormalized in the same way. These results are the same as the
results valid for the undeformed Wess-Zumino model. 
However, unlike in the undeformed Wess-Zumino model, 
divergences in the three-point and four-point functions 
appear. The divergences appearing in the three-point 
functions have their counterparts in the classical action while the 
counterparts for the divergences of the four-point function do not exist in 
the classical action. The five-point function is convergent.

Our results show that in general case with arbitrary coefficients 
$a_1,\dots, a_5$ the model is not renormalizable. However, there is a special 
choice $a_3=a_4=0$ which renders renormalizability. With this choice we still 
have a nontrivial deformation due to the $a_1$-term and the $a_5$-term in the 
action. The divergent part of the four-point function vanishes and the 
two-point and three-point functions are renormalizable. In the special 
case $a_5=-\frac{1}{2}ma_1$ the divergent part of the three-point function 
also vanishes. Equations (\ref{a_1ren}) and (\ref{a_5ren}) are consistent 
with this choice and results (\ref{redfNew}) resemble the results from the 
undeformed Wess-Zumino model.

Having in mind results of \cite{on-shell}, we also investigated on-shell  
renormalizability of the general model. In general, on-shell renormalizability 
leads to a one-loop renormalizable $S$-matrix. On the other hand, one-loop
renormalizable Green functions spoil renormalizability
at higher loops. After using the equations of motion to obtain on-shell 
divergent terms in our model, we are left with some non-vanishing 
five-point divergences proportional to $a_3$ and $a_4$. Since the 
five-point Green function is convergent, these five-point divergences cannot 
be canceled unless $a_3=a_4=0$. But we already know that in that case the model is fully renormalizable.

Renormalization of the deformed Wess-Zumino model has been
previously studied in the literature, see for example
\cite{1/2WZrenorm}, \cite{NACYM}. The models considered there are
non-hermitian and have half of the supersymmetry of the
corresponding ${\cal N}=1$ theory. In general they are not
power-counting renormalizable, however it has been argued in
\cite{1/2WZrenorm}, \cite{NACYM} that they are nevertheless
renormalizable since only a finite number of additional terms needs
to be added to the action to absorb divergences to all orders.
This fact is related to the non-hermiticity of the relevant
actions. Renormalizability of the deformed Wess-Zumino model in
\cite{1/2WZrenorm} was achieved by adding a finite number of
additional terms to the original classical action and using the
equations of motion to eliminate auxiliary field $F$ from the
theory.

The models \cite{1/2WZrenorm}, \cite{NACYM} are different from the
model we study here. The main difference lies in the deformation,
resulting in a different $\star$-product and a different deformed
action. The action of our model is hermitian\footnote{The
deformation (\ref{star}) is not hermitian. However, adding the
complex conjugate terms by hand we obtain the hermitian action
(\ref{S-cl-kin})-(\ref{S-cl-int}).} and moreover, invariant under
the full ${\cal N} = 1$ SUSY. Since our model is more complicated then those
discussed in \cite{1/2WZrenorm} it is not obvious which terms
should be added in the general case with arbitrary $a_3$ nd $a_4$ in order to cancel all one-loop divergences, and
whether the number of the added terms is finite. This remains to
be investigated in future.

Another problem which we plan to address is the choice of deformation. Namely, renormalizability can be chosen as a criterion to test the deformation. We could chose a different deformation compared to that discussed in this paper. Using our principles (SUSY invariance, commutative limit, minimal deformation) we could construct an invariant action and check whether the obtained model has a better behavior. This could give us an important insight into which deformation of the superspace is preferred.

\vspace*{0.5cm}
\begin{flushleft}
{\Large {\bf Acknowledgments}}
\end{flushleft}

The work of the authors is supported by the
project $141036$ of the Serbian Ministry of Science. M.D. thanks
INFN Gruppo collegato di Alessandria for their financial support
during her stay in Alessandria, Italy where a part of this
work was completed.  We also thank Carlos Tamarit for useful comments on 
one-loop on-shell renormalizability and Maja Buri\' c for her useful comments 
on the paper.

\newpage

\appendix
\section{Calculation of traces}

In this appendix we give calculations of some of the traces
appearing in $\Gamma^{(3)}_1$ and $\Gamma^{(4)}_1$. Up to second
order in the deformation parameter the trace $\tr
(AFB\bar{F}\bar{B}F)$ is given by
\begin{eqnarray}
& &\tr (AFB\bar{F}\bar{B}F)= \nonumber\\
& &\hspace{0.5cm}=\int \prod_{i=1}^{4}{\mbox{d}}^8
z_{i}\hspace{1mm} \Big(
\frac{m(1-a_{1}\bar{C}^{2}\square)D^{2}}{4\square (\square-
m^{2})} \Big)_{1}\delta(z_{1}-z_{2})\Big(-\frac{\lambda}{2}\Phi
\bar{D}^{2}\Big)_{2} \nonumber \\ & &\hspace{1cm} \cdot
\Big(\frac{D^{2}\bar{D}^2}{16\square (\square -m^{2})} \Big)_{2}
\delta(z_{2}-z_{3}) \Big( -\frac{\lambda}{2}\Phi^{+} D^{2}
\Big)_{3} \nonumber \\ & &\hspace{1cm} \cdot
\Big(\frac{\bar{D}^{2}D^{2}}{16\square (\square- m^{2})} \Big)_{3}
\delta(z_{3}-z_{4})\Big(-\frac{\lambda}{2}\Phi
\bar{D}^{2}\Big)_{4}\delta(z_{4}-z_{1}) \nonumber \\
& &\hspace{0.5cm}+ \int \prod_{i=1}^{4}{\mbox{d}}^8
z_{i}\hspace{1mm} \Big( \frac{mD^{2}}{4\square (\square-m^{2})}
\Big)_{1}\delta(z_{1}-z_{2})\Big(-\frac{\lambda}{2}\Phi
\bar{D}^{2}\Big)_{2}\nonumber \\ & &\hspace{1cm} \cdot
\Big(\frac{D^{2}\bar{D}^2}{16\square (\square-m^{2})} \Big)_{2}
\delta(z_{2}-z_{3}) \Big( -\frac{\lambda}{2}\Phi^{+} D^{2}
\Big)_{3} \nonumber \\ & &\hspace{1cm} \cdot
\Big(\frac{\bar{D}^{2}D^{2}}{16\square (\square-m^{2})} \Big)_{3}
\delta(z_{3}-z_{4})\nonumber \\ & &\hspace{1cm} \cdot
\Big[\overleftarrow{\bar{D}^{2}}\Big(-\frac{\lambda
C^{2}}{128}(a_{3}D^{2}\Phi- 2a_{4}\Phi^{+})\Big)
D^{2}\bar{D}^{2}\Big]_{4}\delta(z_{4}-z_{1}) \nonumber \\
& &\hspace{0.5cm}+\int \prod_{i=1}^{4}{\mbox{d}}^8
z_{i}\hspace{1mm} \Big( \frac{mD^{2}}{4\square (\square- m^{2})}
\Big)_{1}\delta(z_{1}-z_{2})\Big(-\frac{\lambda}{2}\Phi
\bar{D}^{2}\Big)_{2} \nonumber \\ & &\hspace{1cm} \cdot
\Big(\frac{D^{2}\bar{D}^2}{16\square (\square -m^{2})} \Big)_{2}
\delta(z_{2}-z_{3}) \Big( -\frac{\lambda}{2}\Phi^{+} D^{2}
\Big)_{3} \nonumber \\ & &\hspace{1cm} \cdot
\Big(\frac{\bar{D}^{2}D^{2}}{16\square (\square- m^{2})} \Big)_{3}
\delta(z_{3}-z_{4})\Big(\frac{\lambda a_{5}\bar{C}^2}{8}
(\bar{D}^2 \Phi^{+})
\bar{D}^{2}\Big)_{4} \delta(z_{4}-z_{1}) \nonumber \\
& &\hspace{0.5cm}+ \int \prod_{i=1}^{4}{\mbox{d}}^8
z_{i}\hspace{1mm} \Big( \frac{mD^{2}}{4\square (\square-m^{2})}
\Big)_{1}\delta(z_{1}-z_{2})\Big(-\frac{\lambda}{2}\Phi
\bar{D}^{2}\Big)_{2} \nonumber \\ & &\hspace{1cm} \cdot
\Big(\frac{D^{2}\bar{D}^2}{16\square (\square-m^{2})} \Big)_{2}
\delta(z_{2}-z_{3}) \Big[\overleftarrow{D^{2}}\Big(-\frac{\lambda
\bar{C}^{2}}{128}(a_{3}\bar{D}^{2}\Phi^{+}- 2a_{4}\Phi)\Big)
\bar{D}^{2}D^{2}\Big]_{3} \nonumber \\ & &\hspace{1cm}
\cdot\Big(\frac{\bar{D}^{2}D^{2}}{16\square (\square-m^{2})}
\Big)_{3} \delta(z_{3}-z_{4}) \Big(-\frac{\lambda}{2}\Phi
\bar{D}^{2}\Big)_{4}\delta(z_{4}-z_{1}) \nonumber \\
& &\hspace{0.5cm}+\int \prod_{i=1}^{4}{\mbox{d}}^8
z_{i}\hspace{1mm} \Big( \frac{m D^{2}}{4\square (\square- m^{2})}
\Big)_{1}\delta(z_{1}-z_{2})\Big(-\frac{\lambda}{2}\Phi
\bar{D}^{2}\Big)_{2} \nonumber \\ & &\hspace{1cm} \cdot
\Big(\frac{D^{2}\bar{D}^2}{16\square (\square -m^{2})} \Big)_{2}
\delta(z_{2}-z_{3}) \Big( \frac{\lambda a_{5} C^{2}}{8}(D^2\Phi)
D^{2} \Big)_{3} \nonumber \\ & &\hspace{1cm} \cdot
\Big(\frac{\bar{D}^{2}D^{2}}{16\square (\square- m^{2})} \Big)_{3}
\delta(z_{3}-z_{4})\Big(-\frac{\lambda}{2}\Phi
\bar{D}^{2}\Big)_{4}\delta(z_{4}-z_{1}) \nonumber \\
& &\hspace{0.5cm}+ \int \prod_{i=1}^{4}{\mbox{d}}^8
z_{i}\hspace{1mm} \Big( \frac{mD^{2}}{4\square (\square-m^{2})}
\Big)_{1}\delta(z_{1}-z_{2})\nonumber \\ & &\hspace{1cm} \cdot
\Big[\overleftarrow{\bar{D}^{2}}\Big(-\frac{\lambda
C^{2}}{128}(a_{3}D^{2}\Phi- 2a_{4}\Phi^{+})\Big)
D^{2}\bar{D}^{2}\Big]_{2} \Big(\frac{D^{2}\bar{D}^2}{16\square
(\square-m^{2})} \Big)_{2} \delta(z_{2}-z_{3}) \nonumber \\ &
&\hspace{1cm} \cdot \Big(-\frac{\lambda}{2}\Phi^{+} D^{2}\Big)_{3}
\Big(\frac{\bar{D}^{2}D^{2}}{16\square (\square-m^{2})} \Big)_{3}
\delta(z_{3}-z_{4}) \Big(-\frac{\lambda}{2}\Phi
\bar{D}^{2}\Big)_{4}\delta(z_{4}-z_{1}) \nonumber \\
& &\hspace{0.5cm}+\int \prod_{i=1}^{4}{\mbox{d}}^8
z_{i}\hspace{1mm} \Big( \frac{m D^{2}}{4\square (\square- m^{2})}
\Big)_{1}\delta(z_{1}-z_{2})\Big(\frac{\lambda a_{5}
\bar{C}^2}{8}(\bar{D}^2 \Phi^{+}) \bar{D}^{2}\Big)_{2} \nonumber \\
& &\hspace{1cm} \cdot \Big(\frac{D^{2}\bar{D}^2}{16\square
(\square -m^{2})} \Big)_{2} \delta(z_{2}-z_{3}) \Big(
-\frac{\lambda}{2}\Phi^{+} D^{2} \Big)_{3} \nonumber \\ &
&\hspace{1cm} \cdot \Big(\frac{\bar{D}^{2}D^{2}}{16\square
(\square- m^{2})} \Big)_{3}
\delta(z_{3}-z_{4})\Big(-\frac{\lambda}{2}\Phi
\bar{D}^{2}\Big)_{4}\delta(z_{4}-z_{1}) \nonumber \ .
\end{eqnarray}
After applying the $D$-algebra identities, we obtain
\begin{eqnarray}
& &\tr (AFB\bar{F}\bar{B}F)= \nonumber\\
& &\hspace{0.5cm}= \int {\mbox{d}}^4 \theta \hspace{1mm} \int
\prod_{i=1}^{3}{\mbox{d}}^4 x_{i}\hspace{1mm}
(-8m\lambda^{3})\Big(\frac{1-a_{1}\bar{C}^{2}\square}{\square -
m^{2}}\Big)_{1} \delta(x_{1}-x_{2}) \Phi(x_{2},\theta) \nonumber
\\ & &\hspace{1cm} \cdot \Big(\frac{1}{\square - m^{2}}\Big)_{2} \delta(x_{2}-x_{3})
\Phi^{+}(x_{3}, \theta) \Big(\frac{1}{\square- m^{2}}\Big)_{3}
\delta(x_{3}-x_{1}) \Phi(x_{1}, \theta) \nonumber \\
& &\hspace{0.5cm}+ \int {\mbox{d}}^4 \theta \hspace{1mm} \int
\prod_{i=1}^{3}{\mbox{d}}^4 x_{i}\hspace{1mm}
(-2m\lambda^{3}C^{2}) \Big(\frac{1}{\square-m^{2}}\Big)_{1}
\delta(x_{1}-x_{2}) \Phi(x_{2}, \theta) \nonumber \\ &
&\hspace{1cm} \cdot \Big(\frac{1}{\square-m^{2}}\Big)_{2}
\delta(x_{2}-x_{3}) \Phi^{+}(x_{3}, \theta)
\Big(\frac{\square}{\square-m^{2}}\Big)_{3} \delta(x_{3}-x_{1})
\nonumber \\ & &\hspace{1cm} \cdot \Big(a_{3} (D^{2}\Phi)(x_{1},
\theta)- 2a_{4} \Phi^{+}(x_{1}, \theta)\Big) \nonumber \\
& &\hspace{0.5cm}+ \int {\mbox{d}}^4 \theta \hspace{1mm} \int
\prod_{i=1}^{3}{\mbox{d}}^4 x_{i}\hspace{1mm} (2m\lambda^{3} a_{5}
\bar{C}^{2})\Big(\frac{1}{\square - m^{2}}\Big)_{1}
\delta(x_{1}-x_{2}) \Phi(x_{2},\theta) \nonumber
\\ & &\hspace{1cm} \cdot \Big(\frac{1}{\square - m^{2}}\Big)_{2} \delta(x_{2}-x_{3})
\Phi^{+}(x_{3}, \theta) \Big(\frac{1}{\square- m^{2}}\Big)_{3}
\delta(x_{3}-x_{1}) (\bar{D}^{2}\Phi^{+})(x_{1}, \theta) \nonumber \\
& &\hspace{0.5cm}+ \int {\mbox{d}}^4 \theta \hspace{1mm} \int
\prod_{i=1}^{3}{\mbox{d}}^4 x_{i}\hspace{1mm}
(-2m\lambda^{3}\bar{C}^{2}) \Big(\frac{1}{\square-m^{2}}\Big)_{1}
\delta(x_{1}-x_{2}) \Phi(x_{2}, \theta) \nonumber \\ &
&\hspace{1cm} \cdot \Big(\frac{1}{\square-m^{2}}\Big)_{2}
\delta(x_{2}-x_{3}) \Big(a_{3} (\bar{D}^{2}\Phi^{+})(x_{3},
\theta)- 2a_{4} \Phi(x_{3}, \theta)\Big) \nonumber \\ &
&\hspace{1cm} \cdot \Big(\frac{\square}{\square-m^{2}}\Big)_{3}
\delta(x_{3}-x_{1})
\Phi(x_{1}, \theta) \nonumber \\
& &\hspace{0.5cm}+ \int {\mbox{d}}^4 \theta \hspace{1mm} \int
\prod_{i=1}^{3}{\mbox{d}}^4 x_{i}\hspace{1mm} (2m\lambda^{3} a_{5}
C^{2})\Big(\frac{1}{\square - m^{2}}\Big)_{1} \delta(x_{1}-x_{2})
\Phi(x_{2},\theta) \nonumber
\\ & &\hspace{1cm} \cdot \Big(\frac{1}{\square - m^{2}}\Big)_{2} \delta(x_{2}-x_{3})
(D^{2}\Phi)(x_{3}, \theta) \Big(\frac{1}{\square- m^{2}}\Big)_{3}
\delta(x_{3}-x_{1}) \Phi(x_{1}, \theta) \nonumber \\
& &\hspace{0.5cm}+ \int {\mbox{d}}^4 \theta \hspace{1mm} \int
\prod_{i=1}^{3}{\mbox{d}}^4 x_{i}\hspace{1mm}
(-2m\lambda^{3}C^{2}) \Big(\frac{1}{\square-m^{2}}\Big)_{1}
\delta(x_{1}-x_{2})  \nonumber \\ & &\hspace{1cm} \cdot \Big(a_{3}
(D^{2}\Phi)(x_{2}, \theta)- 2a_{4} \Phi^{+}(x_{2}, \theta)\Big)
\Big(\frac{\square}{\square-m^{2}}\Big)_{2} \delta(x_{2}-x_{3})
\nonumber \\ & &\hspace{1cm} \cdot \Phi^{+}(x_{3}, \theta)
\Big(\frac{1}{\square-m^{2}}\Big)_{3} \delta(x_{3}-x_{1})
\Phi(x_{1}, \theta) \nonumber \\
& &\hspace{0.5cm}+ \int {\mbox{d}}^4 \theta \hspace{1mm} \int
\prod_{i=1}^{3}{\mbox{d}}^4 x_{i}\hspace{1mm} (2m\lambda^{3} a_{5}
\bar{C}^{2})\Big(\frac{1}{\square - m^{2}}\Big)_{1}
\delta(x_{1}-x_{2}) (\bar{D}^{2}\Phi^{+})(x_{2},\theta) \nonumber
\\ & &\hspace{1cm} \cdot \Big(\frac{1}{\square - m^{2}}\Big)_{2} \delta(x_{2}-x_{3})
\Phi^{+}(x_{3}, \theta) \Big(\frac{1}{\square- m^{2}}\Big)_{3}
\delta(x_{3}-x_{1}) \Phi(x_{1}, \theta)\ .\label{eq: AFBbarFbarBF}
\end{eqnarray}
Transforming the previous expression to the momentum space and
using the dimensional regularization we obtain (\ref{TrG3'}).

Similarly, we find
\begin{eqnarray}
& &\tr (B\bar{G}B\bar{F}\bar{B}F)= \nonumber\\
& &\hspace{0.5cm}=\int \prod_{i=1}^{4}{\mbox{d}}^8
z_{i}\hspace{1mm} \Big(\frac{D^{2}\bar{D}^{2}}{16 \square
(\square- m^{2})}\Big)_{1} \delta(z_{1}-z_{2}) \Big[
\overleftarrow{D^2} \Big(\frac{\lambda a_{4}}{64}(\bar{C}^{2}
\bar{D}^{2}\Phi^{+} + C^{2} D^{2} \Phi) \Big) \bar{D}^{2}
\Big]_{2} \nonumber \\ & &\hspace{1cm} \cdot
\Big(\frac{D^{2}\bar{D}^{2}}{16 \square (\square- m^{2})}\Big)_{2}
\delta(z_{2}-z_{3}) \Big( -\frac{\lambda}{2} \Phi^{+} D^{2}
\Big)_{3} \nonumber \\ & &\hspace{1cm} \cdot
\Big(\frac{\bar{D}^{2}D^{2}}{16 \square (\square- m^{2})}\Big)_{3}
\delta(z_{3}-z_{4}) \Big( -\frac{\lambda}{2} \Phi \bar{D}^{2}
\Big)_{4} \delta(z_{4}-z_{1}) \nonumber \\
& &\hspace{0.5cm}+\int \prod_{i=1}^{4}{\mbox{d}}^8
z_{i}\hspace{1mm} \Big(\frac{D^{2}\bar{D}^{2}}{16 \square
(\square- m^{2})}\Big)_{1} \delta(z_{1}-z_{2}) \Big[
\overleftarrow{D^2} \Big(\frac{\lambda a_{5}}{8}(\bar{C}^{2} \Phi
+ C^{2} \Phi^{+}) \Big) \bar{D}^{2} \Big]_{2} \nonumber
\\ & &\hspace{1cm} \cdot \Big(\frac{D^{2}\bar{D}^{2}}{16 \square
(\square- m^{2})}\Big)_{2} \delta(z_{2}-z_{3}) \Big(
-\frac{\lambda}{2} \Phi^{+} D^{2} \Big)_{3} \nonumber \\ &
&\hspace{1cm} \cdot \Big(\frac{\bar{D}^{2}D^{2}}{16 \square
(\square- m^{2})}\Big)_{3} \delta(z_{3}-z_{4}) \Big(
-\frac{\lambda}{2} \Phi \bar{D}^{2}
\Big)_{4} \delta(z_{4}-z_{1}) \nonumber \\
& &\hspace{0.5cm}= \int {\mbox{d}}^4 \theta \hspace{1mm} \int
\prod_{i=1}^{3}{\mbox{d}}^4 x_{i}\hspace{1mm}
(\lambda^{3}a_{4}\bar{C}^{2}) \Big(\frac{\square}{\square - m^{2}}
\Big)_{1} \delta(x_{1}-x_{2}) (\bar{D}^{2}\Phi^{+})(x_{2}, \theta)
\nonumber \\ & &\hspace{1cm} \cdot \Big(\frac{1}{\square- m^{2}}
\Big)_{2} \delta(x_{2}-x_{3}) \Phi^{+}(x_{3}, \theta)
\Big(\frac{1}{\square -m^{2}}\Big)_{3} \delta(x_{3}-x_{1})
\Phi(x_{1}, \theta) \nonumber \\
& &\hspace{0.5cm}+ \int {\mbox{d}}^4 \theta \hspace{1mm} \int
\prod_{i=1}^{3}{\mbox{d}}^4 x_{i}\hspace{1mm}
(\lambda^{3}a_{4}C^{2}) \Big(\frac{1}{\square - m^{2}} \Big)_{1}
\delta(x_{1}-x_{2}) (D^{2}\Phi)(x_{2}, \theta) \nonumber
\\ & &\hspace{1cm} \cdot \Big(\frac{\square}{\square- m^{2}} \Big)_{2}
\delta(x_{2}-x_{3}) \Phi^{+}(x_{3}, \theta) \Big(\frac{1}{\square
-m^{2}}\Big)_{3} \delta(x_{3}-x_{1}) \Phi(x_{1}, \theta) \nonumber\\
& &\hspace{0.5cm}+ \int {\mbox{d}}^4 \theta \hspace{1mm} \int
\prod_{i=1}^{3}{\mbox{d}}^4 x_{i}\hspace{1mm}
(8\lambda^{3}a_{5}\bar{C}^{2}) \Big(\frac{\square}{\square -
m^{2}} \Big)_{1} \delta(x_{1}-x_{2}) \Phi (x_{2}, \theta) \nonumber \\
& &\hspace{1cm} \cdot \Big(\frac{1}{\square- m^{2}} \Big)_{2}
\delta(x_{2}-x_{3}) \Phi^{+}(x_{3}, \theta) \Big(\frac{1}{\square
-m^{2}}\Big)_{3} \delta(x_{3}-x_{1})
\Phi(x_{1}, \theta) \nonumber \\
& &\hspace{0.5cm}+ \int {\mbox{d}}^4 \theta \hspace{1mm} \int
\prod_{i=1}^{3}{\mbox{d}}^4 x_{i}\hspace{1mm}
(8\lambda^{3}a_{5}C^{2}) \Big(\frac{1}{\square - m^{2}} \Big)_{1}
\delta(x_{1}-x_{2}) \Phi^{+}(x_{2}, \theta) \nonumber
\\ & &\hspace{1cm} \cdot \Big(\frac{\square}{\square- m^{2}} \Big)_{2}
\delta(x_{2}-x_{3}) \Phi^{+}(x_{3}, \theta) \Big(\frac{1}{\square
-m^{2}}\Big)_{3} \delta(x_{3}-x_{1}) \Phi(x_{1}, \theta) \ .
\label{eq: BbarGBbarFbarBF}
\end{eqnarray}
The divergent part of (\ref{eq: BbarGBbarFbarBF}) is (\ref{TrG3''}).

The term contributing to the divergent part of the four-point Green function is
\begin{eqnarray}
& &\tr (B\bar{F}\bar{B}FB\bar{F}\bar{B}F)= \nonumber\\
& &\hspace{0.5cm}=\int \prod_{i=1}^{5}{\mbox{d}}^8
z_{i}\hspace{1mm} \Big( \frac{D^{2}\bar{D}^{2}}{16\square
(\square- m^{2})} \Big)_{1} \delta(z_{1}-z_{2})
\Big(-\frac{\lambda}{2}\Phi^{+} D^{2}\Big)_{2} \nonumber \\
& &\hspace{1cm} \cdot \Big(\frac{\bar{D}^2D^{2}}{16\square
(\square -m^{2})} \Big)_{2} \delta(z_{2}-z_{3}) \Big(
-\frac{\lambda}{2}\Phi \bar{D}^{2} \Big)_{3} \nonumber \\ &
&\hspace{1cm} \cdot \Big( \frac{D^{2}\bar{D}^{2}}{16\square
(\square- m^{2})} \Big)_{3} \delta(z_{3}-z_{4})
\Big(-\frac{\lambda}{2}
\Phi^{+} D^{2}\Big)_{4} \nonumber \\
& &\hspace{1cm} \cdot \Big(\frac{\bar{D}^{2} D^{2}}{16 \square
(\square - m^{2})} \Big)_{4} \delta(z_{4}-z_{5}) \Big(-
\frac{\lambda}{2} \Phi \bar{D}^{2} \Big)_{5} \delta(z_{5}- z_{1}) \nonumber\\
& &\hspace{0.5cm}+ 2 \int \prod_{i=1}^{5}{\mbox{d}}^8
z_{i}\hspace{1mm} \Big( \frac{D^{2}\bar{D}^{2}}{16\square
(\square- m^{2})} \Big)_{1} \delta(z_{1}-z_{2}) \nonumber \\
& &\hspace{1cm} \cdot \Big[\overleftarrow{D^{2}} \Big(-
\frac{\lambda \bar{C}^{2}}{128} (a_{3} \bar{D}^{2}\Phi^{+} - 2
a_{4} \Phi)\Big) \bar{D}^{2} D^{2} \Big]_{2} \nonumber \\
& &\hspace{1cm} \cdot \Big(\frac{\bar{D}^2D^{2}}{16\square
(\square -m^{2})} \Big)_{2} \delta(z_{2}-z_{3}) \Big(
-\frac{\lambda}{2}\Phi \bar{D}^{2} \Big)_{3} \Big(
\frac{D^{2}\bar{D}^{2}}{16\square (\square- m^{2})} \Big)_{3}
\delta(z_{3}-z_{4}) \nonumber \\
& &\hspace{1cm} \cdot \Big(-\frac{\lambda}{2} \Phi^{+}
D^{2}\Big)_{4}  \Big(\frac{\bar{D}^{2} D^{2}}{16 \square (\square
- m^{2})} \Big)_{4} \delta(z_{4}-z_{5}) \Big(- \frac{\lambda}{2}
\Phi \bar{D}^{2} \Big)_{5} \delta(z_{5}- z_{1}) \nonumber\\
& &\hspace{0.5cm}+ 2 \int \prod_{i=1}^{5}{\mbox{d}}^8
z_{i}\hspace{1mm} \Big( \frac{D^{2}\bar{D}^{2}}{16\square
(\square- m^{2})} \Big)_{1} \delta(z_{1}-z_{2}) \Big(
\frac{\lambda a_{5} C^{2}}{8}
(D^{2}\Phi) D^{2} \Big)_{2} \nonumber \\
& &\hspace{1cm} \cdot \Big(\frac{\bar{D}^2D^{2}}{16\square
(\square -m^{2})} \Big)_{2} \delta(z_{2}-z_{3}) \Big(
-\frac{\lambda}{2}\Phi \bar{D}^{2} \Big)_{3} \Big(
\frac{D^{2}\bar{D}^{2}}{16\square (\square- m^{2})} \Big)_{3}
\delta(z_{3}-z_{4}) \nonumber \\
& &\hspace{1cm} \cdot \Big(-\frac{\lambda}{2} \Phi^{+}
D^{2}\Big)_{4}  \Big(\frac{\bar{D}^{2} D^{2}}{16 \square (\square
- m^{2})} \Big)_{4} \delta(z_{4}-z_{5}) \Big(- \frac{\lambda}{2}
\Phi \bar{D}^{2} \Big)_{5} \delta(z_{5}- z_{1}) \nonumber\\
& &\hspace{0.5cm}+ 2 \int \prod_{i=1}^{5}{\mbox{d}}^8
z_{i}\hspace{1mm} \Big( \frac{\bar{D}^{2}D^{2}}{16\square
(\square- m^{2})} \Big)_{1} \delta(z_{1}-z_{2}) \nonumber \\
& &\hspace{1cm} \cdot \Big[\overleftarrow{\bar{D}^{2}} \Big(-
\frac{\lambda C^{2}}{128} (a_{3} D^{2}\Phi - 2
a_{4} \Phi^{+})\Big) D^{2} \bar{D}^{2} \Big]_{2} \nonumber \\
& &\hspace{1cm} \cdot \Big(\frac{D^{2}\bar{D}^2}{16\square
(\square -m^{2})} \Big)_{2} \delta(z_{2}-z_{3}) \Big(
-\frac{\lambda}{2}\Phi^{+} D^{2} \Big)_{3} \Big(
\frac{\bar{D}^{2}D^{2}}{16\square (\square- m^{2})} \Big)_{3}
\delta(z_{3}-z_{4}) \nonumber \\
& &\hspace{1cm} \cdot \Big(-\frac{\lambda}{2} \Phi
\bar{D}^{2}\Big)_{4} \Big(\frac{D^{2}\bar{D}^{2}}{16 \square
(\square - m^{2})} \Big)_{4} \delta(z_{4}-z_{5}) \Big(-
\frac{\lambda}{2} \Phi^{+} D^{2} \Big)_{5} \delta(z_{5}- z_{1}) \nonumber\\
& &\hspace{0.5cm}+ 2 \int \prod_{i=1}^{5}{\mbox{d}}^8
z_{i}\hspace{1mm} \Big( \frac{\bar{D}^{2}D^{2}}{16\square
(\square- m^{2})} \Big)_{1} \delta(z_{1}-z_{2}) \Big(\frac{\lambda
a_{5} \bar{C}^{2}}{8}(\bar{D}^{2}\Phi^{+}) \bar{D}^{2} \Big)_{2} \nonumber \\
& &\hspace{1cm} \cdot \Big(\frac{D^{2}\bar{D}^2}{16\square
(\square -m^{2})} \Big)_{2} \delta(z_{2}-z_{3}) \Big(
-\frac{\lambda}{2}\Phi^{+} D^{2} \Big)_{3} \Big(
\frac{\bar{D}^{2}D^{2}}{16\square (\square- m^{2})} \Big)_{3}
\delta(z_{3}-z_{4}) \nonumber \\
& &\hspace{1cm} \cdot \Big(-\frac{\lambda}{2} \Phi
\bar{D}^{2}\Big)_{4} \Big(\frac{D^{2}\bar{D}^{2}}{16 \square
(\square - m^{2})} \Big)_{4} \delta(z_{4}-z_{5}) \Big(-
\frac{\lambda}{2} \Phi^{+} D^{2} \Big)_{5} \delta(z_{5}- z_{1}) \nonumber\\
& &\hspace{0.5cm}= \int \prod_{i=1}^{4}{\mbox{d}}^8
z_{i}\hspace{1mm} (16 \lambda^{4}) \Big(\frac{1}{\square - m^{2}}
\Big)_{1} e^{i(\theta_{1}\sigma^{n}\bar{\theta}_{1} +
\theta_{2}\sigma^{n}\bar{\theta}_{2} -
2\theta_{1}\sigma^{n}\bar{\theta}_{2})\partial_{n}^{1}}
\delta(x_{1}-x_{2})  \nonumber \\
& &\hspace{1cm} \cdot \Phi^{+}(z_{2}) \Big(\frac{1}{\square
-m^{2}} \Big)_{2} \delta(z_{2}-z_{3}) \Phi(z_{3})  \nonumber \\
& &\hspace{1cm} \cdot \Big(\frac{1}{\square - m^{2}} \Big)_{3}
e^{i(\theta_{3}\sigma^{n}\bar{\theta}_{3} +
\theta_{4}\sigma^{n}\bar{\theta}_{4} -
2\theta_{3}\sigma^{n}\bar{\theta}_{4})\partial_{n}^{3}}
\delta(x_{3}-x_{4}) \Phi^{+}(z_{4})  \nonumber \\
& &\hspace{1cm} \cdot  \Big(\frac{1}{\square -m^{2}} \Big)_{4}
\delta(z_{4}-z_{1}) \Phi(z_{1}) \nonumber\\
& &\hspace{0.5cm}+ \int {\mbox{d}}^4 \theta \hspace{1mm} \int
\prod_{i=1}^{4}{\mbox{d}}^4 x_{i}\hspace{1mm} (8 \lambda^{4}
\bar{C}^{2}) \Big( \frac{\square}{\square -m^{2}}\Big)_{1}
\delta(x_{1}-x_{2})  \nonumber \\ & &\hspace{1cm} \cdot (a_{3}
(\bar{D}^{2} \Phi^{+})(x_{2}, \theta) - 2a_{4} \Phi(x_{2},
\theta)) \Big( \frac{\square}{\square -m^{2}}\Big)_{2}
\delta(x_{2}-x_{3}) \Phi(x_{3},\theta) \nonumber \\ &
&\hspace{1cm} \cdot \Big( \frac{1}{\square -m^{2}} \Big)_{3}
\delta(x_{3}-x_{4}) \Phi^{+}(x_{4}, \theta) \Big( \frac{1}{\square
- m^{2}} \Big)_{4}
\delta (x_{4}-x_{1}) \Phi(x_{1}, \theta) \nonumber\\
& &\hspace{0.5cm}+ \int \prod_{i=1}^{4}{\mbox{d}}^8
z_{i}\hspace{1mm} (-8 \lambda^{4} a_{5} C^{2})
\Big(\frac{1}{\square - m^{2}} \Big)_{1}
e^{i(\theta_{1}\sigma^{n}\bar{\theta}_{1} +
\theta_{2}\sigma^{n}\bar{\theta}_{2} -
2\theta_{1}\sigma^{n}\bar{\theta}_{2})\partial_{n}^{1}}
\delta(x_{1}-x_{2})  \nonumber \\
& &\hspace{1cm} \cdot (D^{2}\Phi)(z_{2}) \Big(\frac{1}{\square
-m^{2}} \Big)_{2} \delta(z_{2}-z_{3}) \Phi(z_{3})  \nonumber \\
& &\hspace{1cm} \cdot \Big(\frac{1}{\square - m^{2}} \Big)_{3}
e^{i(\theta_{3}\sigma^{n}\bar{\theta}_{3} +
\theta_{4}\sigma^{n}\bar{\theta}_{4} -
2\theta_{3}\sigma^{n}\bar{\theta}_{4})\partial_{n}^{3}}
\delta(x_{3}-x_{4}) \Phi^{+}(z_{4})  \nonumber \\
& &\hspace{1cm} \cdot  \Big(\frac{1}{\square -m^{2}} \Big)_{4}
\delta(z_{4}-z_{1}) \Phi(z_{1}) \nonumber\\
& &\hspace{0.5cm}+ \int {\mbox{d}}^4 \theta \hspace{1mm} \int
\prod_{i=1}^{4}{\mbox{d}}^4 x_{i}\hspace{1mm} (8 \lambda^{4}
C^{2}) \Big( \frac{\square}{\square -m^{2}}\Big)_{1}
\delta(x_{1}-x_{2})  \nonumber \\ & &\hspace{1cm} \cdot (a_{3}
(D^{2} \Phi)(x_{2}, \theta) - 2a_{4} \Phi^{+}(x_{2}, \theta))
\Big( \frac{\square}{\square -m^{2}}\Big)_{2} \delta(x_{2}-x_{3})
\Phi^{+}(x_{3},\theta) \nonumber \\ & &\hspace{1cm} \cdot \Big(
\frac{1}{\square -m^{2}} \Big)_{3} \delta(x_{3}-x_{4}) \Phi(x_{4},
\theta) \Big( \frac{1}{\square - m^{2}} \Big)_{4} \delta
(x_{4}-x_{1}) \Phi^{+}(x_{1}, \theta) \nonumber\\
& &\hspace{0.5cm}+ \int \prod_{i=1}^{4}{\mbox{d}}^8
z_{i}\hspace{1mm} (-8 \lambda^{4} a_{5} \bar{C}^{2})
\Big(\frac{1}{\square - m^{2}} \Big)_{1}
e^{-i(\theta_{1}\sigma^{n}\bar{\theta}_{1} +
\theta_{2}\sigma^{n}\bar{\theta}_{2} -
2\theta_{2}\sigma^{n}\bar{\theta}_{1})\partial_{n}^{1}}
\delta(x_{1}-x_{2})  \nonumber \\
& &\hspace{1cm} \cdot (\bar{D}^{2}\Phi^{+})(z_{2})
\Big(\frac{1}{\square
-m^{2}} \Big)_{2} \delta(z_{2}-z_{3}) \Phi^{+}(z_{3})  \nonumber \\
& &\hspace{1cm} \cdot \Big(\frac{1}{\square - m^{2}} \Big)_{3}
e^{-i(\theta_{3}\sigma^{n}\bar{\theta}_{3} +
\theta_{4}\sigma^{n}\bar{\theta}_{4} -
2\theta_{4}\sigma^{n}\bar{\theta}_{3})\partial_{n}^{3}}
\delta(x_{3}-x_{4}) \Phi(z_{4})  \nonumber \\
& &\hspace{1cm} \cdot  \Big(\frac{1}{\square -m^{2}} \Big)_{4}
\delta(z_{4}-z_{1}) \Phi^{+}(z_{1}) \ . \label{eq:
BbarFbarBFBbarFbarBF}
\end{eqnarray}
From (\ref{eq: BbarFbarBFBbarFbarBF}) the divergence (\ref{TrG4}) follows.

All divergences appearing in (\ref{eq: AFBbarFbarBF}), (\ref{eq:
BbarGBbarFbarBF}) and (\ref{eq: BbarFbarBFBbarFbarBF}) are
obtained using the following formulae:
\begin{eqnarray}
&&\int \prod_{i=1}^{3}{\mbox{d}}^4 x_{i}\hspace{1mm}
\Big(\frac{\square}{\square-m^{2}}\Big)_{1} \delta(x_{1}-x_{2})
f_{1}(x_{2}) \Big(\frac{1}{\square-m^{2}}\Big)_{2}
\delta(x_{2}-x_{3}) f_{2}(x_{3}) \nonumber \\
&& \hspace{6.5cm}
\cdot \Big(\frac{1}{\square-m^{2}}\Big)_{3} \delta(x_{3}-x_{1})
f_{3}(x_{1})\Big|_{dp} \nonumber\\
&&\hspace{3cm}= \frac{i}{8\pi^{2}\varepsilon} \int
{\mbox{d}}^4 x\hspace{1mm} f_{1}(x)f_{2}(x)f_{3}(x) ,\label{app1}
\end{eqnarray}
\begin{eqnarray}
&&\int \prod_{i=1}^{4}{\mbox{d}}^4 x_{i}\hspace{1mm}
\Big(\frac{\square}{\square-m^{2}}\Big)_{1} \delta(x_{1}-x_{2})
f_{1}(x_{2}) \Big(\frac{\square}{\square-m^{2}}\Big)_{2}
\delta(x_{2}-x_{3}) f_{2}(x_{3}) \nonumber \\
&& \hspace{1.7cm}
\cdot \Big(\frac{1}{\square-m^{2}}\Big)_{3} \delta(x_{3}-x_{4})
f_{3}(x_{4}) \Big(\frac{1}{\square-m^{2}}\Big)_{4}
\delta(x_{4}-x_{1}) f_{4}(x_{1})\Big|_{dp} \nonumber \\
&&\hspace{3cm}  = \frac{i}{8\pi^{2}\varepsilon} \int {\mbox{d}}^4
x\hspace{1mm} f_{1}(x)f_{2}(x)f_{3}(x) f_{4}(x) .\label{app2}
\end{eqnarray}


\begin{thebibliography}{99}

\bibitem{luksusy}
P.~Kosi{\' n}ski, J.~Lukierski and P.~Ma{\' s}lanka, {\it Quantum
Deformations of Space-Time SUSY and Noncommutative Superfield
Theory}, hep-th/0011053.

P.~Kosi{\' n}ski, J.~Lukierski, P.~Ma{\' s}lanka and J.~Sobczyk, {\it
Quantum Deformation of the Poincare Supergroup and
$\kappa$-deformed Superspace}, J.\ Phys. A {\bf 27} (1994) 6827,
[hep-th/9405076].

\bibitem{MWsusy}
Chong-Sun Chu and F.~Zamora, {\it Manifest supersymmetry in noncommutative geometry}, JHEP {\bf 0002}, 022 (2000), [hep-th/9912153].

S.~Ferrara and M.~A.~Lledo, {\it Some aspects of deformations of
supersymmetric field theories}, JHEP {\bf 05}, 008 (2000), [hep-th/0002084].

\bibitem{nonantisusy}
D.~Klemm, S.~Penati and L.~Tamassia, {\it Non(anti)commutative
superspace}, Class.\ Quant.\ Grav. {\bf 20} (2003) 2905,
[hep-th/0104190].

J.~de Boer, P.~A.~Grassi and P.~van Nieuwenhuizen, {\it
Noncommutative superspace from string theory}, Phys. Lett. B {\bf
574}, 98 (2003), [hep-th/0302078].

\bibitem{Seiberg}
N.~Seiberg, {\it Noncommutative superspace, $N = 1/2$
supersymmetry, field theory and string theory}, JHEP {\bf 0306}
010 (2003), [hep-th/0305248].

\bibitem{Ferrara}
S. Ferrara, M. Lledo and O. Macia, {\it Supersymmetry in
noncommutative superspaces}, JHEP {\bf 09} (2003) 068, [hep-th/0307039].

\bibitem{14}
B.~M.~Zupnik, {\it Twist-deformed supersymmetries in
non-anticommutative super- spaces}, Phys.\ Lett.\ {\bf B 627} 208
(2005), [hep-th/0506043].

M.~Ihl and C.~S\" amann, {\it Drinfeld-twisted supersymmetry and
non-anticommutative superspace}, JHEP {\bf 0601} (2006) 065, [hep-th/0506057].

\bibitem{D-def-us}
M.~Dimitrijevi\' c and V.~Radovanovi\' c, {\it D-deformed Wess-Zumino model and its renormalizability properties}, JHEP {\bf 0904}, 108 2009, 0902.1864[hep-th].

\bibitem{miSUSY}
M.~Dimitrijevi\' c, V.~Radovanovi\' c and J.~Wess, {\it Field Theory on 
Nonanticommutative Superspace}, JHEP {\bf 0712}, 059 (2007), 0710.1746[hep-th].

\bibitem{Martin08} C.~P.~Martin and C.~Tamarit, {\it The Seiberg-Witten map 
and supersymmetry}, JHEP {\bf  0811 }, 087 (2008), 0809.2684[hep-th].

\bibitem{Martin09} C.~P.~Martin and C.~Tamarit, {\it  Noncommutative $N=1$ 
super Yang-Mills, the Seiberg-Witten map and UV divergences}, JHEP {\bf 0911}, 092 (2009), 0907.2437[hep-th].

\bibitem{Drinf}
V.~G.~Drinfel'd, {\it On constant quasiclassical solutions of the Yang-Baxter equations}, Soviet Math. Dokl. {\bf 28}, 667 (1983).

V.~G.~Drinfel'd, {\it Quantum groups}, In Proc.\ Int.\ Cong.\ Math.\ Berkeley, 798 (1986).

\bibitem{TwistSymm}
M.~Chaichian, P.~Kulish, K.~Nishijima and A.~Tureanu, {\it On a 
Lorentz-Invariant Interpretation of Noncommutative Space-Time and Its 
Implications on Noncommutative QFT}, Phys.\ Lett.\  {\bf B 604}, 98 (2004), [hep-th/0408069].

J.~Wess, {\it Deformed Coordinate Spaces; Derivatives}, in Proceedings of 
the BW2003 Workshop, Vrnjacka Banja, Serbia, 2003, [hep-th/0408080].

\bibitem{defgt}
P.~Aschieri, C.~Blohmann, M.~Dimitrijevi\' c, F.~Meyer, P.~Schupp
and J.~Wess, {\it A Gravity Theory on Noncommutative Spaces},
Class.\ Quant.\ Grav. {\bf 22}, 3511-3522 (2005),
[hep-th/0504183].

P.~Aschieri, M. Dimitrijevi\' c, F.~Meyer, S.~Schraml and J.~Wess,
{\it Twisted Gauge Theories}, Lett.\ Math.\ Phys. 78, 61-71 (2006),
[hep-th/0603024].

D.~V.~Vassilevich, {\it Twist to close}, Mod.\ Phys.\ Lett. A {\bf
21}, 1279 (2006), [hep-th/0602185].

\bibitem{chpr}
V.~G.~Drinfel’d, {\it Quasi-Hopf algebras}, Leningrad Math. J. 1, 1419 (1990).

V.~Chari and A.~Pressley, {\it A Guide to Quantum Groups}, Cambridge University
Press, Cambridge (1995).

\bibitem{twist2}
P.~Aschieri, M. Dimitrijevi\' c, F.~Meyer and  J.~Wess, {\it
Noncommutative Geometry and Gravity}, Class.\ Quant.\ Grav. {\bf
23}, 1883-1912 (2006), [hep-th/0510059].

\bibitem{wessbook}
J.~Wess and J.~Bagger, {\it Supersymmetry and Supergravity},
Princton, USA: Univ.~Pr. (1992).

\bibitem{sw}
S. Weinberg, {\it Quantum Field Theory II}, Cambridge Univesity
Press, New York (1996).

\bibitem{MV}
M.~Buri\' c and V.~Radovanovi\' c, {\it On divergent $3$-vertices in 
noncommutative $SU(2)$ gauge theory}, Class.\ Quant.\ Grav.\ {\bf 22}, 525 
(2005), [hep-th/0410085].

\bibitem{on-shell}
C.~Tamarit, {\it Noncommutative GUT inspired theories with $U(1)$, $SU(N)$ 
groups and their renormalisability}, 0910.5195[hep-th].

G.~’t Hooft and M.~J.~G.~Veltman, {\it One loop divergencies in the theory of 
gravitation}, Annales Poincare Phys.\ Theor.\ {\bf A20}, 69 (1974).

R.~E.~Kallosh, O.~V.~Tarasov and I.~V.~Tyutin, {\it One Loop Finiteness Of 
Quantum Gravity Off Mass Shell}, Nucl.\ Phys.\ {\bf B137}, 145 (1978).


\bibitem{1/2WZrenorm}
S.~Terashima and J.~T.~Yee, {\it Comments on Noncommutative Superspace}, JHEP 
{\bf 0312}, 053 (2003), [hep-th/0306237].

M.~T.~Grisaru, S.~Penati and A.~Romagnoni, {\it Two-loop Renormalization for 
Nonanticommutative $N=1/2$ Supersymmetric WZ Model}, JHEP {\bf 0308}, 003 
(2003), [hep-th/ 0307099].

I.~Jack, D.~R.~T.~Jones and R.~Purdy, {\it The non-anticommutative 
supersymmetric Wess-Zumino model}, 0808.0400[hep-th].

R.~Britto and B.~Feng, {\it $N=1/2$ Wess-Zumino model is renormalizable}, 
Phys.\ Rev.\ Lett.\ {\bf 91}, 201601 (2003), [hep-th/0307165].

A.~Romagnoni,
{\it Renormalizability of N = 1/2 Wess-Zumino model in superspace},
JHEP {\bf 0310}, 016 (2003), [hep-th/0307209]. 

R.~Britto, B.~Feng, Soo-Jong Rey, {\it Non(anti)commutative superspace, UV/IR 
mixing and open Wilson lines}, JHEP {\bf 0308}, 001 (2003), [hep-th/0307091].

R.~Britto, B.~Feng, Soo-Jong Rey, {\it Deformed superspace, N = 1/2 
supersymmetry and nonrenormalization theorems}, JHEP {\bf 0307}, 067 (2003), 
[hep-th/0306215].




\bibitem{NACYM}
O.~Lunin and S.~J.~Rey, {\it Renormalizability of Non(anti)commutative Gauge 
Theories with $N =1/2$ Supersymmetry}, JHEP {\bf 0309}, 045 (2003), 
[hep-th/0307275].

I.~Jack, D.~R.~T. Jones and L.~A.~Worthy, {\it One-loop renormalisation of 
$N = 1/2$ supersymmetric gauge theory with a superpotential}, 
Phys.\ Rev.\ {\bf D75}, 045014 (2007), [hep-th/0701096].

D.~Berenstein, S. J. Rey, {\it Wilsonian proof for renormalizability of N=1/2 
supersymmetric field theories}, Phys.\ Rev.\ D68, 121701, (2003), 
[hep-th/0308049].

\end{thebibliography}
\end{document}